\documentclass{an}
\usepackage{graphicx}
\usepackage{times}
\usepackage{fancyhdr}
\begin{document}

\title{Calibration Observations of Fomalhaut with the VLTI}

\author{J. Davis \inst{1,2}\fnmsep\thanks{Participant in the ESO Visitor
Programme}
    \and A.~Richichi \inst{2}
    \and P.~Ballester \inst{2}
    \and Ph.~Gitton \inst{2}
    \and A.~Glindemann \inst{2}
    \and S.~Morel \inst{2}
    \and M.~Schoeller \inst{2}
    \and M.~Wittkowski \inst{2}
    \and F.~Paresce \inst{2}}
\institute{School of Physics, University of Sydney, NSW 2006, Australia
    \and European Southern Observatory, Garching b. M\"{u}nchen, Germany}

\date{Received; accepted; published online}

\abstract{An investigation of the stability of the transfer
function of the European Southern Observatory's Very Large
Telescope Interferometer has been carried out through observations
of Fomalhaut, which was observed over a range in hour angle from
21:50--05:24 on 20 October 2002.  No significant variation in the
transfer function was found for the zenith angle range
5\degr--70\degr. The projected baseline varied between 139.7~m and
49.8~m during the observations and, as an integral part of the
determination of the transfer function, a new accurate
limb-darkened angular diameter for Fomalhaut of 2.109 $\pm$ 0.013
mas has been established. This has led to improved values for the
emergent flux = (3.43 $\pm$ 0.10)$\times10^{8}$~Wm$^{-2}$,
effective temperature = 8819 $\pm$ 67~K and radius = (1.213 $\pm$
0.011)$\times10^{9}$~m (R/R$_{\sun}$ = 1.744 $\pm$ 0.016). The
luminosity has been found to be
(6.34 $\pm$ 0.20)$\times10^{27}$~W (L/L$_{\sun}$ =
16.5 $\pm$ 0.5).
}

\correspondence{davis@physics.usyd.edu.au}

\maketitle

\section{Introduction}\label{sec:intro}
A major problem in ground-based optical/infrared long-baseline
interferometry is establishing the true interference fringe
visibility from measurements made through the Earth's turbulent
atmosphere.  The observed visibility is always less than the true
visibility due to residual seeing and instrumental effects no
matter how well the interferometer is designed, compensated and
aligned.  The technique generally adopted to correct for the loss
in visibility is to interleave observations of the object being
studied with observations of a calibrator - a star for which the
true visibility at the baseline and wavelength being employed is
known.  A calibrator is thus a star that is either essentially
unresolved or whose angular diameter is known with sufficient
accuracy to predict the true visibility with the desired accuracy.
Division of the observed visibility for the programme object by
the observed visibility for the calibrator yields the true
visibility for the programme object, providing the instrumental
and atmospheric conditions do not change significantly between the
measurements.

Optical/infrared interferometers generally measure the square of
the fringe visibility as defined by Michelson and, for
convenience, we will follow Davis et al. (\cite{susi2}) and use
correlation $C$ to represent the square of the fringe visibility
$V$. We define the transfer function $T_\mathrm{F}$ of an
interferometer as the ratio of the observed correlation
$C_\mathrm{obs}$ over the expected correlation $C_\mathrm{exp}$.
The transfer function is thus given by

\begin{equation}
T_\mathrm{F} = \frac{V_\mathrm{obs}^{2}}{V_\mathrm{exp}^{2}} =
\frac{C_\mathrm{obs}}{C_\mathrm{exp}}. \label{eqn:transfer}
\end{equation}

The true correlation for a programme object is then given by
dividing its observed correlation by the transfer function for its
associated calibrator.

This procedure works well when programme object and calibrator are
close in the sky and the observations are made close in time.
However, problems have been experienced at blue wavelengths
($\sim$440 nm) with the Sydney University Stellar Interferometer
(SUSI) (Davis et al. \cite{susi1}) when the program object and
calibrator are separated by more than $\approx5\degr$ on the sky,
or when the seeing conditions are changing on the timescale of the
individual measurements.

As part of the commissioning programme of the European Southern
Observatory's Very Large Telescope Interferometer (VLTI)
(Glindemann et al. \cite{vlti}), a large number of
calibrator observations have been accumulated
(Percheron et al. \cite{calibvlti}) and
are being pursued to assess, among other things, the stability
and repeatability of the transfer function in the spectral K band
at a wavelength of 2.2~$\mu$m.  The correlation
measurements are made with the VINCI instrument (Kervella et al.
\cite{vinci}), which is based on the FLUOR instrument (Coud\'{e}
du Foresto et al. \cite{fluor}), and it employs an optical fibre
beam-combiner and photometric monitoring of the beams.  The
spatial and temporal scales of the turbulence-induced wavefront
distortions scale as $\lambda^{6/5}$ and, with the spatial
filtering provided by VINCI, the VLTI is expected to be
significantly less affected by atmospheric effects than
instruments working at shorter wavelengths and with conventional
beam-combiners.

One step proposed for the investigative programme was to observe a
single star over a large range in hour angle, and hence in zenith
angle $Z$, to investigate whether the transfer function had a
dependence on $Z$. Time was allocated for these programmes when
the VLTI was operating with its two test siderostats (Derie et al.
\cite{sids}) separated by 140 m. The limiting K magnitude of the
system for an unresolved star was $\sim+3.0$ and, for a partially
resolved star, the limiting magnitude was brighter by $2.5\log
C_\mathrm{exp}$. With this sensitivity limit a shorter baseline
would have provided a choice of unresolved stars for the
experiment. However, given the circumstances, it was decided to
make observations at the longer baseline since they were of
intrinsic interest and, in addition, would provide experience for
planning more detailed experiments with a shorter baseline in the
future. The limiting magnitude placed severe restrictions on the
choice of stars, particularly as an unresolved star with
declination close to the latitude of the VLTI location on Paranal
($\approx-25\degr$) would ideally be chosen for following over a
large range in hour angle. At the scheduled time for the
observations no suitable unresolved candidate stars were
available.  It was decided to follow Fomalhaut ($\alpha$ PsA, HR
8728, HD216956, spectral type A3V), even though it would be
significantly resolved at 140~m based on the Narrabri Stellar
Intensity Interferometer limb-darkened angular diameter of
$(2.10\pm0.14)$~mas (Hanbury Brown et al. \cite{hbdanda}).

\section{The Observations and Data Analysis}\label{sec:obs}
The observations were made on the night of 20 October 2002 in the
normal observing mode of the VLTI as a series of Observing Blocks
(OBs) except that, instead of alternating OBs between a programme
star and a calibrator, all OBs were for Fomalhaut.  Each block
consisted of three groups of 500 scans of the fringe envelope. The
star was followed over the hour angle range from 21:50 to 05:24
giving a range in zenith angle from $5\degr$ to $>70\degr$. The
sky was clear throughout the observations and the mean zenith
seeing, measured in the visual, was 0.76\arcsec.  The seeing only
exceeded 1\arcsec for the first 3 of the 34 accepted sets of
scans.  The atmospheric coherence time, again determined for the
visual, had 99\% of the 502 values between 1.1 ms and 2.7 ms with
a mean value of 1.85 ms.  The range of values for the seeing and
coherence time represent a typical night at Paranal and in no
sense are they exceptional and, as discussed in
Sect.~\ref{sec:TransFunc}, there is no correlation between the
variations in the seeing or coherence time and in the observed
correlation values.

The observational data were processed in the VINCI pipeline
(Ballester et al. \cite{pipeline}).  The processing automatically
excludes scans for which the signal/noise in the photometry
channels is less than 5.  Some 25\% of the sets of scans retained
only a small fraction of the total of 500 scans due to drifting of
the starlight off the input end of the optical fibers. The groups
retaining less than 40\% of the total scans (15 out of a total of
49 sets) were found to deviate randomly from the general trend of
the observations and to have large uncertainties associated with
them. They were removed from the analysis reported here, although
a separate analysis showed that they had no significant influence
on the results or conclusions because of their low statistical
weight.

\begin{figure}
  \resizebox{\hsize}{!}{\includegraphics{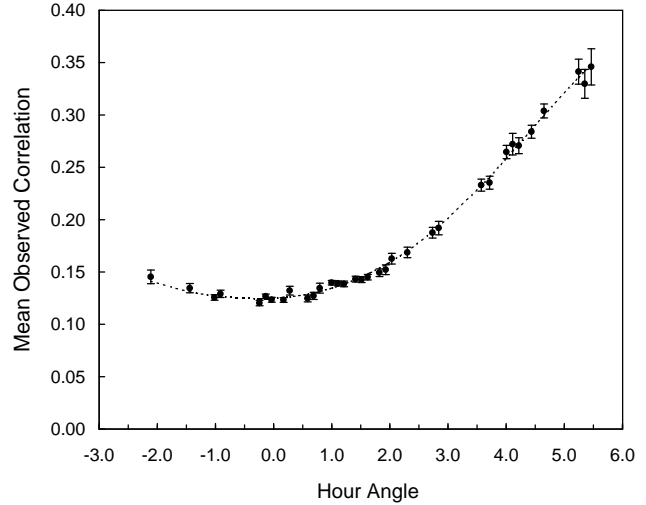}}
  \caption{The correlation measurements as a function of hour angle
  on the night of 20 October 2002.  The broken line is the expected
  correlation for a uniform-disk angular diameter of 2.086 mas
  (see Sect.~\ref{sec:TransFunc}).  Further details are given in the text.}
  \label{fig:channels}
\end{figure}

The correlation values from VINCI's two interferometry channels, A
and B, differ by a scale factor but track each other closely.
Initially they were analysed independently and gave essentially
identical results with no significant differences.  The two sets
of correlation values were therefore put on the same scale by
normalising the B values by the weighted mean ratio of the A/B
values. The two sets of data were then combined by taking a
weighted mean of the A and scaled B values at each hour angle. The
weighted mean values of correlation as a function of hour angle,
zenith angle and projected baseline are listed in
Table~\ref{tab:data}. Figure~\ref{fig:channels} shows the
correlation values plotted against hour angle for the combined
data. The large variation of correlation with hour angle was
expected due to the change in projected baseline from 139.7~m to
49.8~m during the course of the observations.

\begin{table}
  \caption{The weighted mean values of correlation from the two VLTI signal
channels as
  a function of hour angle, zenith angle and projected baseline.  Note that
the formal
  uncertainties  in the correlation values have been increased by a factor
of 2.2 as
  explained in Section~\ref{sec:TransFunc}.}
  \label{tab:data}
  \begin{tabular}{cccr}
  \hline
  Hour   & Projected & Zenith & \multicolumn{1}{c}{Correlation}  \\
  Angle  & Baseline & Angle & \\
  (hours) & (m)  &  (degrees) &           \\
  \hline
21.89 & 132.5 & 28.57 & 0.1453 $\pm$ 0.0065 \\
22.56 & 136.7 & 19.85 & 0.1345 $\pm$ 0.0044 \\
22.98 & 138.4 & 14 48 & 0.1256 $\pm$ 0.0026 \\
23.09 & 138.8 & 13.16 & 0.1291 $\pm$ 0.0034 \\
23.76 & 139.7 &  5.96 & 0.1203 $\pm$ 0.0026 \\
23.86 & 139.7 &  5.32 & 0.1267 $\pm$ 0.0022 \\
23.97 & 139.6 &  5.01 & 0.1234 $\pm$ 0.0023 \\
00.17 & 139.2 &  5.48 & 0.1233 $\pm$ 0.0022 \\
00.28 & 138.9 &  6.21 & 0.1321 $\pm$ 0.0042 \\
00.58 & 137.8 &  9.23 & 0.1247 $\pm$ 0.0032 \\
00.69 & 137.2 & 10.44 & 0.1271 $\pm$ 0.0034 \\
00.79 & 136.6 & 11.69 & 0.1345 $\pm$ 0.0047 \\
01.00 & 135.3 & 14.19 & 0.1397 $\pm$ 0.0022 \\
01.10 & 134.5 & 15.53 & 0.1391 $\pm$ 0.0023 \\
01.21 & 133.7 & 16.87 & 0.1385 $\pm$ 0.0027 \\
01.41 & 131.9 & 19.40 & 0.1432 $\pm$ 0.0028 \\
01.51 & 130.8 & 20.75 & 0.1427 $\pm$ 0.0028 \\
01.62 & 129.7 & 22.12 & 0.1450 $\pm$ 0.0026 \\
01.82 & 127.4 & 24.73 & 0.1493 $\pm$ 0.0037 \\
01.93 & 126.0 & 26.12 & 0.1522 $\pm$ 0.0047 \\
02.03 & 124.6 & 27.50 & 0.1628 $\pm$ 0.0051 \\
02.30 & 120.8 & 30.96 & 0.1687 $\pm$ 0.0051 \\
02.73 & 113.6 & 36.64 & 0.1875 $\pm$ 0.0052 \\
02.84 & 111.7 & 38.02 & 0.1919 $\pm$ 0.0064 \\
03.57 &  96.8 & 47.50 & 0.2330 $\pm$ 0.0058 \\
03.71 &  93.5 & 49.33 & 0.2352 $\pm$ 0.0062 \\
04.01 &  86.6 & 53.09 & 0.2646 $\pm$ 0.0064 \\
04.11 &  83.9 & 54.46 & 0.2721 $\pm$ 0.0103 \\
04.22 &  81.2 & 55.82 & 0.2706 $\pm$ 0.0076 \\
04.44 &  75.6 & 58.59 & 0.2840 $\pm$ 0.0062 \\
04.65 &  69.8 & 61.32 & 0.3038 $\pm$ 0.0067 \\
05.25 &  53.0 & 68.79 & 0.3413 $\pm$ 0.0119 \\
05.35 &  49.8 & 70.12 & 0.3297 $\pm$ 0.0138 \\
05.46 &  46.5 & 71.51 & 0.3460 $\pm$ 0.0173 \\
  \hline
  \end{tabular}
\end{table}

\section{The Effective Wavelength} \label{sec:wavelength}
In interferometric measurements, the spatial frequency
of each observation is given by the ratio of the baseline
and the wavelength. While the baseline can be reliably
measured and its projection easily predicted as a function
of observation time, an accurate determination of
the wavelength requires some effort since the measurements
are usually performed using filters with a finite bandwidth.
The spectral sensitivity curve of the VLTI VINCI system has been
determined from the spectral transmission curve of the K band
filter used, the attenuation curve for the optical fibers, the
detector spectral quantum efficiency curve, and representative
transmission curves for the atmosphere across the band (Manduca \&
Bell \cite{mandb}; Lord \cite{lord}). The spectral reflectivities
of the silver and gold coated mirrors in the optical path are
essentially constant across the K band. The spectral sensitivity
curves were computed for the two atmospheric transmission curves
and gave effective wavelengths, for a constant photon flux per
unit wavelength across the spectrum, of 2.199~$\mu$m and
2.195~$\mu$m.  The mean of these two values, 2.197~$\mu$m, has
been adopted.  Based on the differing model predictions for the
atmospheric transmission across the band, the range of
precipitable water vapour at Paranal, and the effect of zenith
angle, the uncertainty in the effective wavelength is estimated to
be $\pm$0.005~$\mu$m.
More details on this approach can be found in a technical
report by Davis \& Richichi (\cite{effwav}).
For the observations of Fomalhaut, taking
into account the slope of the stellar spectrum across the spectral
sensitivity curve, the effective wavelength was determined to be
2.181 $\pm$ 0.005~$\mu$m. This value is insensitive to the
temperature of the star and changing the temperature by
$\pm$~2000~K changes the effective wavelength by $<$0.001~$\mu$m.

\section{The Transfer Function}\label{sec:TransFunc}
The variation in correlation seen in
Fig.~\ref{fig:channels} is primarily due to the variation in
projected baseline during the course of the observations and
follows closely the curve expected for a partially resolved star.
If there is a variation in the transfer function $T_\mathrm{F}$
with time this will also contribute to the variation of
correlation but to a very much smaller degree.

Initially $T_\mathrm{F}$ was computed as a function of
zenith angle with the expected correlation $C_\mathrm{exp}$
computed using the Intensity Interferometer angular diameter.  The
limb-darkened angular diameter of 2.10 mas (Hanbury Brown et al.,
\cite{hbdanda}) was converted to the equivalent uniform-disk
angular diameter for the K band for this purpose using a
correction factor ($\rho$) taken from Davis et al. (\cite{ld1})
who computed correction factors for Kurucz model atmospheres.  For
an A3V star, assuming an effective temperature of 8750~K and $\log
g$ of 4.0, the ratio of the limb-darkened to uniform-disk angular
diameter ($\rho$) is 1.011 giving an equivalent uniform-disk
angular diameter of 2.08 mas for the K band.  The computed
values of $T_\mathrm{F}$ were almost constant with zenith angle
showing only a very small linear slope. However, the
distribution of the observational points about the line of best
fit had a greater spread than expected for the formal
uncertainties in the individual values. It was clear that
the uncertainties were underestimated and,
although the cause cannot be identified from the data, short term
effects due to the atmosphere are a likely significant
contributor.  The formal uncertainties were increased by a scaling
factor and $\chi^{2}$ was computed for a linear fit to the
$T_\mathrm{F}$ versus zenith angle data. This was repeated for a
range of scaling factors and it was found that a scaling factor of
2.2 gave a $\chi^{2}$ probability of $\sim$70\%. This factor has
been adopted and applied to the formal uncertainties. The scaled
uncertainties have been used throughout the following analysis and
discussion. It is important to stress that this has no effect on
the results or conclusions of this paper apart from increasing the
uncertainties in the derived parameters.

Although $T_\mathrm{F}$ is almost constant with zenith
angle based on the angular diameter determined with the NSII, a
variation of $T_\mathrm{F}$ with zenith angle cannot be excluded.
If a variation is present it will affect the angular diameter
determined from the values of correlation versus projected
baseline. Assuming for the moment that $T_\mathrm{F}$ is constant
with zenith angle, and following the general practice for stars
with compact atmospheres, the values of correlation as a function
of projected baseline were fitted with the curve for a uniform
disk

\begin{equation}
C_\mathrm{exp} = \left|\frac{2J_\mathrm{1}(x)}{x}\right|^{2}
\label{eqn:udisk}
\end{equation}

\noindent where $x = \pi b \theta_\mathrm{UD}/\lambda$, $b$ is the
projected baseline of the interferometer, $\theta_\mathrm{UD}$ is
the equivalent uniform-disk angular diameter of the star and
$\lambda$ is the effective wavelength of the observations.  The
projected baselines were computed in the VLTI data-processing
pipeline and the effective wavelength for the observations was
2.181 $\pm$ 0.005 $\mu$m (Sect.~\ref{sec:wavelength}). The
fit yielded a uniform-disk angular diameter of  2.086$\pm$
0.012~mas and a correlation value at zero baseline of 0.3904
$\pm$ 0.0048 with a reduced $\chi^{2}$ value of 0.83. The
uncertainty in the angular diameter is the formal error of the
fit.

\begin{figure}
  \resizebox{\hsize}{!}{\includegraphics{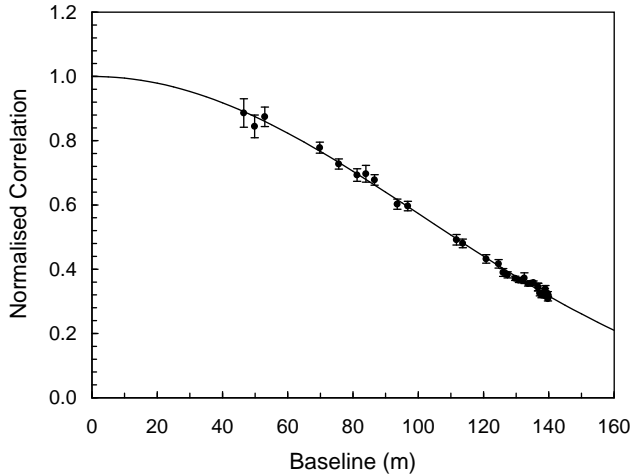}}
  \caption{Normalised observed correlation values versus projected baseline.
  The curve is the fit for the equivalent uniform-disk angular diameter of
2.086 mas.}
  \label{fig:transform}
\end{figure}

When the uncertainty in the effective wavelength
(Sect.~\ref{sec:wavelength}) is taken into account, the
uniform-disk angular diameter becomes 2.086 $\pm$ 0.013~mas.
The observed correlation values, normalised by the zero-baseline
correlation value, and the fitted equivalent uniform disk curve
are shown in Fig.~\ref{fig:transform}.

It can be seen from Fig.~\ref{fig:transform} that the fit is
excellent with no obvious trends away from the uniform disk curve.
The points at the shortest baselines, which show the greatest
scatter and largest uncertainties, were taken at zenith angles
exceeding 65\degr but have negligible effect on the resulting
uniform-disk angular diameter.  There is no correlation between
the differences of the observed correlation from the fitted curve
and the variations in the seeing and atmospheric coherence time.

The transfer function has been assumed constant in the
analysis. If it had varied significantly during the course of the
observations it would have affected the correlation values and led
to an erroneous value for the angular diameter. Assume for the
moment that $T_\mathrm{F}$ varied with zenith angle and that the
variation is known and used to `correct' the observed values of
correlation. An angular diameter fit to the `corrected' values of
correlation would be expected to result in an improved fit with a
lower value for the reduced $\chi^{2}$.  To explore this
possibility various linear dependencies of the $T_\mathrm{F}$ on
both the zenith angle $Z$ and on $\sec^{-3/5}Z$ were tried.  The
latter is the theoretical dependence of the atmospheric spatial
and temporal coherence parameters $r_{0}$ and $t_{0}$ on $Z$.  In
each case the correlation values were `corrected' and an angular
diameter fit was made.  It was found that the minimum value for
the reduced $\chi^{2}$ occurred for zero dependence on $Z$ and on
$\sec^{-3/5}Z$.  It is concluded that any dependence of
$T_\mathrm{F}$ on $Z$ is small and that the angular diameter
determined on the assumption of zero dependence is valid.

\begin{figure}[b]
  \resizebox{\hsize}{!}{\includegraphics{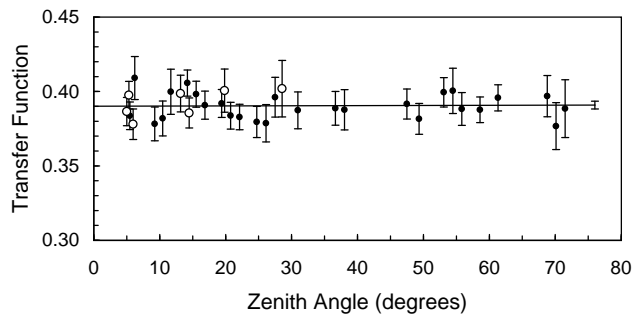}}
  \caption{The transfer function $T_\mathrm{F}$ as a function of
  zenith angle.  The line is a linear regression fit to the data.
  Further details are given in the text.}
  \label{fig:tf}
\end{figure}

Finally, $T_\mathrm{F}$ was computed as a function of $Z$,
using the new uniform-disk angular diameter to predict the
expected correlation in equation~\ref{eqn:transfer}. The resulting
values for $T_\mathrm{F}$ are plotted against zenith angle in
Fig.~\ref{fig:tf}.  The figure shows no apparent dependence of
transfer function on zenith angle over the range from
5\degr--70\degr, as confirmed through a linear regression fit.
Again there is no correlation
between the differences of the observed correlation from the
fitted curve and the variations in the seeing and atmospheric
coherence time.  The mean transfer function was 0.3901
$\pm$ 0.0026, in agreement with the zero-baseline correlation
value obtained from the angular diameter fit.

\begin{figure}[t]
  \resizebox{\hsize}{!}{\includegraphics{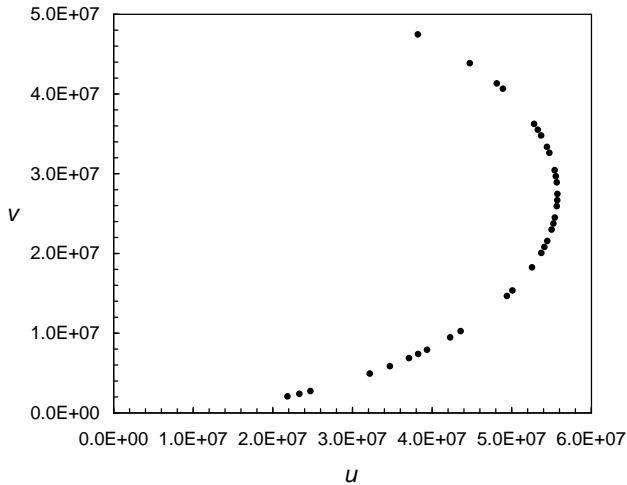}}
  \caption{The {\em u,v} plot of the observations showing the 40\degr range
  in orientation of the projected baseline relative to the star. Further
  details are given in the text.}
  \label{fig:uv}
\end{figure}

It is noted that in determining the angular diameter, circular
symmetry has been assumed.  The value of $V\sin i$ of
85~kms$^{-1}$ for Fomalhaut (Abt \& Morrell, \cite{vsini}) does
not classify it as a rapid rotator, and any effect of rotational
distortion on the observations is small compared with the
uncertainty in the angular diameter. Fig.~\ref{fig:uv} shows the
{\em u,v} track of the observations.  The angle between the
projected baseline and a fixed arbitrary reference axis through
the centre of the star in the plane of the sky changed by only
45.8\degr during the observations.  Fomalhaut is a main-sequence
IR excess star but shows no excess radiation at the wavelength of
observation (Gillett \cite{irexcess}). Scattered radiation from
the star will be a small fraction of the direct stellar radiation
and will not significantly affect the results discussed here. An
independent study based on VLTI measurements of Fomalhaut, using
the standard approach of calibrated observations, also confirms
that there is no detectable infrared excess at the VINCI
wavelengths (Di Folco et al. \cite{difolco}).

The mean transfer function corresponds to a fringe visibility of
62.5\% for a completely unresolved source and it only applies to
the particular night of observation. The transfer function may
vary due to changes in the atmosphere and in the instrument. Long
term changes occur for the VLTI due to a temperature-dependence of
polarization in the fiber beam combiner and changes in the optical
train to configure different baselines.  An investigation of the
long-term changes in the transfer function of the VLTI is being
pursued and values of the fringe visibility between 40\% and 87\%
for an unresolved source have been measured over a period of about
18 months of operation of the VLTI.  The implication of the
present result is that, in the short term and with a single
baseline, the VLTI is remarkably stable.

\section{Fundamental Quantities for Fomalhaut}\label{sec:fundquant}
It was concluded in the previous section that any dependence of
the $T_{F}$ on $Z$ is small and that the uniform disk angular
diameter determined on the assumption of zero dependence is valid.
Based on this assumption, we have proceeded to derive the
limb-darkened angular diameter and some fundamental parameters for
Fomalhaut. The factor $\rho$, for converting the uniform disk
angular diameter to the true limb-darkened angular diameter, has
been computed to be 1.011 $\pm$ 0.001 for the relatively broad
spectral passband of the VLTI following the procedure given by
Tango \& Davis (\cite{ld2}). Thus the true limb-darkened angular
diameter is 2.109 $\pm$ 0.013~mas. This is in excellent agreement
with the limb-darkened angular diameter determined with the
Narrabri Stellar Intensity Interferometer of 2.10 $\pm$ 0.14~mas
and with a factor of $>$10 improvement in accuracy from $\sim\pm
6.7\%$ to $\sim\pm 0.6\%$.

The combination of the new accurate limb-darkened angular diameter
with the flux received from the star, and with the Hipparcos
parallax, enables improved values for the emergent flux at the
stellar surface, and the star's effective temperature and radius
to be calculated.

The flux received from Fomalhaut was derived by Code et al.
(\cite{codeetal}).  Following the same procedures, the integrated
flux has been revised by putting the visual spectrophotometry of
Davis \& Webb (\cite{dandw}) on the Vega calibration of Hayes
(\cite{hayes}) in place of that of Oke \& Schild (\cite {oands}),
and using the absolute flux calibration at 550~nm of M\'{e}gessier
(\cite{meg}).  The infrared contribution has been estimated from
the flux distribution for Fomalhaut given by Alekseeva et al.
(\cite{baltic}), scaled to match the Davis \& Webb flux
distribution in the range 700--800~nm, the JHKL photometry of
Bouchet et al. (\cite{bms}) converted to fluxes using
M\'{e}gessier's calibration (\cite{meg}), and the 12~$\mu$m flux
given by Gillett (\cite{irexcess}).  The ultraviolet contribution
in the band 182--330~nm has been reduced by 2.5\% in accord with
the change in the visual flux between 330~nm and 370~nm to which
the uv spectrophotometry was originally matched by Code et al. The
flux shortward of 182~nm has not been changed. The total flux
received from the star, integrated over the entire spectrum, is
estimated to be (8.96 $\pm$ 0.25)$\times10^{-9}$~Wm$^{-2}$.

Combining the received flux with the limb-darkened angular
diameter gives the emergent surface flux to be (3.43 $\pm$
0.10)$\times10^{8}$~Wm$^{-2}$ and the effective temperature 8819
$\pm$ 67~K.  The accuracy of the effective temperature has been
improved by a factor of five compared with the determination by
Code et al. (\cite{codeetal}) and is now limited by the
uncertainty in the integrated flux rather than by that of the
angular diameter.  The improved accuracy will assist in
identifying stellar atmosphere models with the star.

The Hipparcos parallax (\cite{hip}) of 130.08 $\pm$ 0.92~mas,
combined with the limb-darkened angular diameter, gives the radius
of Fomalhaut as (1.213 $\pm$ 0.011)$\times10^{9}$~m (R/R$_{\sun}$
= 1.744 $\pm$ 0.016).  Previous estimates of the radius have been
based on the Narrabri Stellar Intensity Interferometer angular
diameter which limited the accuracy to $\sim\pm$6.7\% but, with
the new value for the angular diameter, the accuracy has been
improved to $\pm$0.9\%.  The luminosity of Fomalhaut, given by
$4\pi d^{2}f$, where $d$ is the distance and $f$ is the total flux
received from the star, is (6.34 $\pm$ 0.20)$\times10^{27}$~W
(L/L$_{\sun}$ = 16.5 $\pm$ 0.5).

\section{Summary}\label{sec:summary}
We have analyzed a set of VLTI interferometric measurements
obtained on the bright star Fomalhaut using the VLTI.
These measurements were recorded with the main goal
of studying the accuracy and stability of the VLTI equipped
with the VINCI instrument. Unlike standard interferometric
practice, they were not interspersed with independent
calibrator observations.

For the night concerned, while it is not possible to completely
rule out a dependence on zenith angle, we have shown that the
transfer function was remarkably stable even though the
observations were taken over large ranges in zenith angle,
projected baseline, and observed correlation values.  The
observational data are an excellent fit to the transform for a
uniform disk and our analysis shows that, if there is a dependence
on zenith angle, it is at a level that does not significantly
affect the value of the fitted uniform disk angular diameter.

A bonus, and an integral result of the analysis of the
observations, was the accurate determination of the angular
diameter for Fomalhaut.  This has allowed fundamental quantities
for this important star to be determined with significantly
improved accuracy.

The angular diameter has been determined without recourse to
observations of a calibrator.  This has been possible as a result
of observations having been made over a wide range in projected
baseline length combined with the stability of the transfer
function. It is emphasised that the interleaving of observations
of calibrators with those of programme stars will continue to be
the normal mode of operation, since it would not be realistic to
abandon it on the basis of one night's observations.  However, the
implications of the results presented here are that the VLTI in
its current configuration, using the VINCI optical fiber beam
combiner, gives remarkably stable transfer functions in the short
term with a single baseline, and that the angular separation of
target and calibrators is not as crucial a matter as found at
shorter wavelengths with conventional beam-combiners.  It augurs
well for the future of the VLTI and for optical/infrared
interferometry in general.

\acknowledgements The work presented was carried out during a
visit by J. Davis to the European Southern Observatory during
which he worked with members of the VLTI group on calibration
problems.   He is grateful to F. Paresce, S. D'Odorico, and the
ESO Visitor's Programme for the opportunity to participate in the
VLTI programme.  Assistance from P.~Kervella during the
observations and discussions with E.~Di Folco are gratefully
acknowledged. A.~Richichi also wishes to acknowledge the support
received during an extended visit to the School of Physics,
University of Sydney, sponsored by the ESO Director General's
Discretionary Fund.



\begin{thebibliography}{}
\bibitem[1995]{vsini}
  Abt, H. A., \& Morrell, N. I. 1995, ApJS, 99, 135
\bibitem[1997]{baltic}
  Alekseeva, G. A., Arkharov, A. A., Galkin, V. D. et al. 1997,
  Baltic Astron., 6, 481
\bibitem[2002]{pipeline}
  Ballester, P., Chavan, A. M., Cotton, B. et al. 2001, in
  Proc. SPIE Vol. 4477, Astronomical Data Analysis,
  eds. J. L. Starck \& F. Murtagh, 225
\bibitem[1991]{bms}
  Bouchet, P., Manfroid, J., \& Schmider, F. X. 1991, A\&AS, 91,
  409
\bibitem[1976]{codeetal}
  Code, A. D., Davis, J., Bless, R. C., \& Hanbury Brown, R. 1976,
  ApJ, 203, 417/
\bibitem[1998]{fluor}
  Coud\'{e} du Foresto, V., Perrin, G., Ruilier, C., et al. 1998, in
  Proc. SPIE Vol. 3350, Astronomical Interferometry, ed. R.
  Reasenberg, 856
\bibitem[1974]{dandw}
  Davis, J., \& Webb, R. J. 1974, MNRAS, 168, 163
\bibitem[1999a]{susi1}
  Davis, J., Tango, W. J., Booth, A. J., et al. 1999a, MNRAS, 303, 773
\bibitem[1999b]{susi2}
  Davis, J., Tango, W. J., Booth, A. J., Thorvaldson, E. D., \& Giovannis,
J. 1999b,
  MNRAS, 303, 783
\bibitem[2000]{ld1}
  Davis, J., Tango, W. J., \& Booth, A.J. 2000, MNRAS, 318, 387
\bibitem[2003]{effwav}
  Davis, J., Richichi, A.,
2003, ESO Technical Report VLT-TRE-ESO-15810-3033
\bibitem[2000]{sids}
  Derie, F., Brunetto, E., Duchateau, M., et al. 2000, in Proc.
  SPIE Vol. 4006, Interferometry in Optical Astronomy, eds. P. J.
  L\'{e}na, \& A. Qirrenbach, 99
\bibitem[2004]{difolco}
  Di Folco, E., Thevenin, F.,  Kervella, P., Domiciano de Souza, A.,
  Coud\'{e} du Foresto, V., et al., 2004, A\&A submitted
\bibitem[1997]{hip}
  ESA 1997, The Hipparcos Catalogue, ESA SP-1200
\bibitem[1986]{irexcess}
  Gillett, F. C. 1986, in Astrophys. \& Space Sciences Library Vol. 124,
  Proc. of First Iras Conference, Light on Dark Matter, ed. F. P. Israel
  (Reidel, Dordrecht), 61
\bibitem[2000]{vlti}
  Glindemann, A., Abuter, R., Carbognani, F., et al. 2000, in Proc.
  SPIE Vol. 4006, Interferometry in Optical Astronomy, eds. P. J.
  L\'{e}na, \& A. Qirrenbach, 2
\bibitem[1974]{hbdanda}
  Hanbury Brown, R., Davis, J., \& Allen, L. R. 1974, MNRAS, 167,
  121
\bibitem[1985]{hayes}
  Hayes, D. S. 1985, in Proc. IAU Symposium No. 111, Calibration of
  Fundamental Stellar Quantities, ed. D. S. Hayes, L. E. Pasinetti,
  \& A. G. Davis Philip (Reidel: Dordrecht), 225
\bibitem[2000]{vinci}
  Kervella, P., Coud\'{e} du Foresto, V., Glindemann, A., \&
  Hofmann, R. 2000, in Proc. SPIE Vol. 4006, Interferometry in
  Optical Astronomy, eds. P. J. L\'{e}na, \& A. Qirrenbach, 31
\bibitem[1992]{lord}
  Lord, S. D. 1992, NASA Technical Memor. 103957 \\
  (Data available from \\
  www.gemini.edu/sciops/ObsProcess/obsConstraints/)
\bibitem[1979]{mandb}
  Manduca, A., \& Bell, R. A. 1979, PASP, 91, 858
\bibitem[1995]{meg}
  M\'{e}gessier, C. 1995, A\&A, 296, 771
\bibitem[1970]{oands}
  Oke, J. B., \& Schild, R. E. 1970, ApJ, 161, 1015
\bibitem[2003]{calibvlti}
  Percheron, I., Richichi, A., Wittkowski, M., 2003, Astrophysics
and Space Science, 286, 219
\bibitem[2002]{ld2}
  Tango, W. J., \& Davis, J. 2002, MNRAS, 333, 642
\end{thebibliography}
\end{document}